\documentclass[
reprint,
amsmath,amssymb,
aps, pre,
]{revtex4-2}

\usepackage[title]{appendix}
\usepackage{graphicx}
\usepackage{dcolumn}
\usepackage{dsfont}
\usepackage{bm}
\usepackage[hidelinks]{hyperref}
\usepackage[capitalise]{cleveref}
\crefname{appendix}{Appendix}{Appendices}
\Crefname{appendix}{Appendix}{Appendices}
\usepackage{lipsum}
\usepackage[dvipsnames]{xcolor}
\usepackage{amsfonts}
\usepackage{yfonts}
\usepackage{booktabs}
\usepackage{amsthm}
\usepackage[scr=boondoxo,scrscaled=1.05]{mathalfa}

\theoremstyle{definition}

\newcommand{\Cov}{\operatorname{Cov}}
\newcommand{\diag}{\operatorname{diag}}

\newcommand{\Var}{\operatorname{Var}}

\DeclareFontFamily{U}{futm}{}
\DeclareFontShape{U}{futm}{m}{n}{
  <-> s * [.97534] fourier-bb 
  }{}
\DeclareMathAlphabet{\mathbbs}{U}{futm}{m}{n}

\begin{document}

\preprint{APS/123-QED}

\title{Event Order as a Resource for Time Estimation in Experimental Non-Markovian Dynamics}

\author{Jiming Zheng} \email{jiming@unc.edu}

\author{Zhiyue Lu}

\affiliation{Department of Chemistry, University of North Carolina-Chapel Hill, NC}

\date{\today}

\begin{abstract}
Biological timekeeping relies on physical signals that are often subject to substantial fluctuations. Counting random events is a natural way to measure time. In systems with memory, however, a larger event count may reflect either a longer elapsed time or a period of faster dynamics. Can event order help resolve this ambiguity? Guided by the kinetic and clock uncertainty relations for non-Markovian dynamics, we construct two clocks directly from experimental event records: one uses only event counts, while the other also retains event order. Once calibrated, neither clock requires access to event timestamps. In spontaneous retinal activity, the timing advantage of retaining event order persists across observation durations from 0.5 to 16 seconds in both wild-type and mutant mice. In two recordings from cultured neuronal networks, retaining event order reduces timing variance by 60 and 70 percent, respectively. We identify the recurrence of events on high-frequency electrodes as a major source of this improvement and interpret these signals as reflecting the activity of local neuronal populations. Our results demonstrate that memory encoded in event order is an experimentally accessible resource for improving the precision of stochastic clocks.
\end{abstract}

\maketitle

\section{Introduction}
\label{sec:introduction}

The ability to measure time is essential to many biological functions, from the regulation of circadian rhythms to the timing of actions \cite{patke2020molecular,shinomoto2011deciphering}. At microscopic scales, counting random events provides a natural way to estimate elapsed time \cite{barato2016cost}. At a constant mean event rate, the expected number of events grows linearly with elapsed time. Fluctuations in the number of events, however, limit the precision with which time can be inferred.

In many biological systems, the occurrence of an event depends on the preceding history. Examples include fluctuations in enzymatic turnover rates, correlated conformational changes in proteins, dwell-time correlations in ion channels, and history-dependent neuronal firing \cite{lu1998single,english2006ever,vollmar2024model,rothberg1997two,truccolo2005point,farkhooi2009serial}. Such memory can arise when the observed dynamics leave internal degrees of freedom unresolved \cite{hummer2015optimal,hartich2021emergent,zhao2024emergence}. Fluctuations in these hidden degrees of freedom can change the speed of the observed dynamics, making it difficult to distinguish a longer elapsed time from a period of faster dynamics. Ordered event sequences have been used to infer hidden transition pathways \cite{zhao2025towards} and dissipation from coarse-grained observations \cite{blom2024milestoning,schwarz2026consistent}. Can the information carried by event order also help resolve the ambiguity between elapsed time and dynamical speed?

For Markov jump processes, the kinetic uncertainty relation (KUR) and clock uncertainty relation (CUR) establish fundamental precision limits for stochastic clocks \cite{di2019kinetic,macieszczak2024ultimate,prech2025optimal}. In autonomous Markov jump dynamics with resolved transition channels, unordered counts and ordered sequences carry the same timing information, provided the boundary states are retained \cite{prech2025optimal}. Uncertainty relations have also been established for particular classes of non-Markovian dynamics \cite{vu2020generalized,coghi2026levelprl,coghi2026level}. In the companion Letter \cite{zheng2026kinetic}, we derive kinetic and clock uncertainty relations for general non-Markovian jump dynamics, establishing a hierarchy of precision limits for different levels of observation. The theory reveals that memory can break the equivalence between ordered and unordered event records. Here we address a complementary question: how much of this additional timing information can be exploited in experimental records, and which properties of the event sequence make it useful?

We construct two time estimators directly from experimental records: one uses only the channel counts, and the other also uses event order. Neither estimator uses the event timestamps. By comparing the two estimators and examining the effect of randomizing event order at fixed channel counts, we assess the contribution of sequential structure to timing precision. Using recordings of spontaneous retinal activity \cite{stafford2009spatial,eglen2014data,eglen2014dataset}, we show that the timing advantage of event order persists across a broad range of observation durations in both wild-type and mutant mice. We then examine the physical origin of this advantage in cultured neuronal networks \cite{lonardoni2017recurrently,lonardoni2018dataset}. In two recordings, retaining event order reduces timing variance by approximately $60\%$ and $70\%$, respectively. We trace much of this improvement to recurrence statistics and find that high-frequency electrodes contribute most of the timing signal carried by recurrence. Together, these results demonstrate that the timing information encoded in event order can be extracted directly from biological dynamics and exploited to improve the precision of stochastic clocks.

\section{Timing Information and Experimental Clocks}
\label{sec:information}
\label{sec:estimators}

\begin{figure}[t]
  \centering
  \includegraphics[width=\columnwidth]{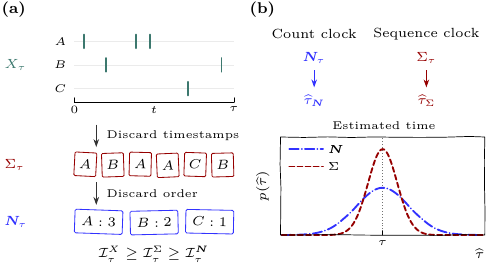}
  \caption{\textbf{Event order as a resource for time estimation.} (a) Three representations of the same event record and their hierarchy of timing information. Discarding timestamps yields the ordered sequence $\Sigma_\tau$, and further discarding event order yields the channel counts $\boldsymbol{N}_\tau$. (b) Two clocks estimating the elapsed time from the counts and the ordered sequence, respectively. The schematic distributions illustrate how retaining event order can reduce timing fluctuations at the same mean readout. The vertical dotted line marks the actual duration $\tau$.}
  \label{fig:observation}
\end{figure}

We consider a stochastic system that generates discrete events in $M$ observed channels, labeled by $e = 1, \ldots, M$. The event rates may depend on the entire preceding history of the system. An event record of duration $\tau$ can be represented at three levels, as illustrated in \cref{fig:observation}(a),
\begin{equation}
  X_\tau \longrightarrow \Sigma_\tau \longrightarrow \boldsymbol{N}_\tau,
  \label{eq:records}
\end{equation}
where $X_\tau = \{(e_k,t_k)\}_{k=1}^{n_\tau}$ retains the event labels and timestamps, $\Sigma_\tau = (e_1, \ldots, e_{n_\tau})$ retains the labels in their original order, and $\boldsymbol{N}_\tau = \{n_e(\tau)\}_e$ retains the count of each label. Here $n_\tau = \sum_e n_e(\tau)$ is the total number of events. When a state trajectory is resolved, the boundary states are also retained at all three levels.

To quantify the timing information contained in each record, we consider a uniform rescaling of the dynamical speed by a factor $c$, such that an event occurring at time $t$ in the original dynamics occurs at time $t/c$ in the accelerated dynamics. This operation rescales all internal time scales, including those associated with memory \cite{zheng2026kinetic}. For a record $\mathcal{O}_\tau \in \{X_\tau,\Sigma_\tau,\boldsymbol{N}_\tau\}$, let $\mathcal{P}_c^{\mathcal{O}}[\mathcal{O}_\tau]$ denote its probability or probability density in the accelerated dynamics. The corresponding score and Fisher information are
\begin{align}
  \Lambda_\tau^{\mathcal{O}} &\equiv \left. \partial_{\ln c} \ln \mathcal{P}_c^{\mathcal{O}}[\mathcal{O}_\tau] \right|_{c=1}, \\
  \mathcal{I}_\tau^{\mathcal{O}} &\equiv \left\langle \left(\Lambda_\tau^{\mathcal{O}}\right)^2 \right\rangle,
\end{align}
where the average is taken over the original dynamics. The Fisher information quantifies how sensitively the statistics of an observed record change with the dynamical speed. Since each level of observation is obtained by discarding information from the preceding one, the scores satisfy $\Lambda_\tau^{\mathcal{O}} = \mathbb{E}[\Lambda_\tau^X \,|\, \mathcal{O}_\tau]$, giving the information hierarchy
\begin{equation}
  \mathcal{I}_\tau^{\boldsymbol{N}} \leq \mathcal{I}_\tau^\Sigma \leq \mathcal{I}_\tau^X.
  \label{eq:information_hierarchy}
\end{equation}
The first gap quantifies the timing information lost when event order is discarded, and the second quantifies that lost when timestamps are discarded. In autonomous Markov jump dynamics with resolved transition channels, $\mathcal{I}_\tau^{\boldsymbol{N}} = \mathcal{I}_\tau^\Sigma$ when the boundary states are retained \cite{prech2025optimal}. Memory can lift this equivalence, making event order an additional source of timing information \cite{zheng2026kinetic}.

This hierarchy motivates the construction of two stochastic clocks, i.e., time estimators, from the same event record, as illustrated in \cref{fig:observation}(b). Both clocks estimate elapsed time without access to the event timestamps, but the sequence clock can additionally exploit event order to suppress timing fluctuations.

For an untagged record, accelerating the autonomous dynamics by a factor $c$ is statistically equivalent to extending the observation duration from $\tau$ to $c\tau$, provided the initial preparation is compatible with time dilation. For a time estimator $\widehat{\tau}_{\mathcal{O}}$ constructed from $\mathcal{O}_\tau \in \{ \boldsymbol{N}_\tau, \Sigma_\tau \}$, whose readout does not explicitly depend on the unknown duration, this equivalence gives
\begin{equation}
  \tau \partial_\tau \langle \widehat{\tau}_{\mathcal{O}} \rangle
  = \Cov[\widehat{\tau}_{\mathcal{O}}, \Lambda_\tau^{\mathcal{O}}].
\end{equation}
The Cauchy--Schwarz inequality then yields the precision bound
\begin{equation}
  J_{\mathcal{O}}(\tau) \equiv \frac{\left[ \tau \partial_\tau \langle \widehat{\tau}_{\mathcal{O}} \rangle \right]^2}{\Var[\widehat{\tau}_{\mathcal{O}}]} \leq \mathcal{I}_\tau^{\mathcal{O}} \leq \mathcal{I}_\tau^X,
  \label{eq:clock_precision}
\end{equation}
where the full-record bound is the non-Markovian KUR, while the sequence and count information give the CUR and the counting bound, respectively. For a clock with unit mean rate, $J_{\mathcal{O}} = 1/\Var[\widehat{\tau}_{\mathcal{O}} / \tau]$ is the inverse relative timing variance. We characterize the improvement of the sequence clock by the ratio
\begin{equation}
  \eta \equiv \frac{J_\Sigma(\tau)}{J_{\boldsymbol{N}}(\tau)} = \left( \frac{s_\Sigma}{s_{\boldsymbol{N}}} \right)^2 \frac{V_{\boldsymbol{N}}}{V_\Sigma},
  \label{eq:order_gain}
\end{equation}
where $s_{\mathcal{O}} \equiv \partial_\tau \langle \widehat{\tau}_{\mathcal{O}} \rangle$ and $V_{\mathcal{O}} \equiv \Var[\widehat{\tau}_{\mathcal{O}}]$. Note that the explicit dependence on $\tau$ cancels in $\eta$, so the timing gain can be evaluated without an absolute time calibration. Here $s_{\mathcal{O}}$ measures the mean clock rate: it equals unity when the mean clock reading advances at the same rate as physical time. The rate factor makes $J_{\mathcal{O}}$ invariant under a nonzero rescaling of the clock reading. Thus $\eta > 1$ indicates an improvement in timing precision.

We construct each clock as a linear combination of statistics of its record,
\begin{equation}
  \widehat{\tau}_{\mathcal{O}} = b_{\mathcal{O}} + \boldsymbol{a}_{\mathcal{O}}^{\mathsf{T}} \boldsymbol{\phi}_{\mathcal{O}}(\mathcal{O}_\tau).
  \label{eq:clock_readout}
\end{equation}
The vector $\boldsymbol{\phi}_{\mathcal{O}} = (\phi_{\mathcal{O},1}, \phi_{\mathcal{O},2}, \ldots)^{\mathsf{T}}$ contains statistics of the event record, and $\boldsymbol{a}_{\mathcal{O}} = (a_{\mathcal{O},1}, a_{\mathcal{O},2}, \ldots)^{\mathsf{T}}$ contains their corresponding coefficients. The scalar $b_{\mathcal{O}}$ is an offset. The count statistics include individual channel counts and nonlinear functions of those counts. The sequence clock uses the same count statistics and adds statistics of event pairs, repeated labels, and the arrangement of events in space. Let $\boldsymbol{r}_e$ denote the spatial position associated with event label $e$. The positions are fixed throughout the recording and available to both clocks. All sequence statistics depend only on the event labels and their order, without using timestamps. For each order statistic, we subtract its mean over uniform permutations of the labels at fixed counts. The statistics used to construct the two estimators are specified in \cref{app:features}.

To calibrate each clock, we sample records of known durations $\tau_j$, $j = 1, \ldots, m$. For each duration, we calculate the mean vector $\boldsymbol{m}_j$ and covariance matrix $C_j$, whose components are
\begin{align}
  (\boldsymbol{m}_j)_k &= \langle \phi_{\mathcal{O},k} \rangle_j, \\
  (C_j)_{kl} &= \Cov_j(\phi_{\mathcal{O},k}, \phi_{\mathcal{O},l}),
\end{align}
where $\langle \cdot \rangle_j$ and $\Cov_j$ denote the sample mean and covariance over the calibration records of duration $\tau_j$, respectively. The coefficients are determined by minimizing the average timing variance subject to the condition that the mean estimate equals the known duration,
\begin{subequations}
\label{eq:clock_optimization}
\begin{align}
  \min_{\boldsymbol{a},b} \quad& \boldsymbol{a}^{\mathsf{T}} \overline{C} \boldsymbol{a} + \rho \boldsymbol{a}^{\mathsf{T}} \diag(\overline{C}) \boldsymbol{a}, \\
  \text{subject to} \quad& b + \boldsymbol{a}^{\mathsf{T}} \boldsymbol{m}_j = \tau_j, \qquad j = 1, \ldots, m,
\end{align}
\end{subequations}
where $\overline{C} = m^{-1}\sum_j C_j$. The optimization is performed separately for each clock, with the subscript $\mathcal{O}$ omitted for brevity. The term with $\rho$ penalizes large coefficients relative to the scale of each statistic, reducing the effect of statistical uncertainty in the calibration data. The coefficients follow from solving linear equations (\cref{app:optimization}). The regularization parameter $\rho$ is chosen to minimize the mean squared timing error on a separate validation set (\cref{app:validation}). After calibration, the same coefficients are used for every duration, and neither estimator uses the unknown duration or the event timestamps.

The sequence estimator can reproduce the count estimator by using the same count coefficients and offset and setting all additional coefficients to zero. At fixed $\rho$ and calibration data, adding sequence statistics therefore cannot increase the minimum of \cref{eq:clock_optimization}.

We divide each experimental recording into five consecutive segments. Three segments are used to determine the coefficients of the estimator for a given $\rho$, one to choose $\rho$ by minimizing the mean squared timing error, and the remaining one to measure timing precision. This procedure is repeated so that each segment is used once to measure timing precision. The resulting time estimates from all five segments are then combined. The mean clock rate $s_{\mathcal{O}}$ is calculated from the mean estimates for windows of durations $(1 \pm 0.2)\tau$ with the same starting points. To obtain confidence intervals at each duration, we repeatedly draw groups of windows with nearby starting times, using the same draws for both clocks and keeping their coefficients fixed. To examine the effect of event order, we randomly reorder the labels within each window without changing the counts and determine the coefficients of the sequence clock again. Further details are given in \cref{app:validation}.

\section{Event Order in Spontaneous Retinal Activity}
\label{sec:retina}

We first examine spontaneous electrical activity in the developing mouse retina, recorded from isolated retinal preparations using a 512-electrode array. The activity arises spontaneously without external stimulation, providing an experimental realization of autonomous stochastic dynamics. During early development, groups of neurons become active and recruit neighboring cells, forming propagating patterns known as retinal waves. We analyze all 29 recordings in the Stafford2009 public subset: 17 wild-type (WT) recordings and 12 recordings from mice lacking the $\beta2$ subunit of the nicotinic acetylcholine receptor ($\beta2$KO). This receptor plays an important role in the generation of retinal waves \cite{stafford2009spatial}. Each event label identifies a recorded neuronal unit, and the dataset provides the two spatial coordinates assigned to that unit \cite{eglen2014data,eglen2014dataset}. We rescale each spatial coordinate to the interval $[0,1]$ before constructing the statistics. The two clocks are calibrated at durations of $0.5$, $1$, $2$, $4$, $8$, and $16\,\mathrm{s}$.

\begin{figure}[t]
  \centering
  \includegraphics[width=\columnwidth]{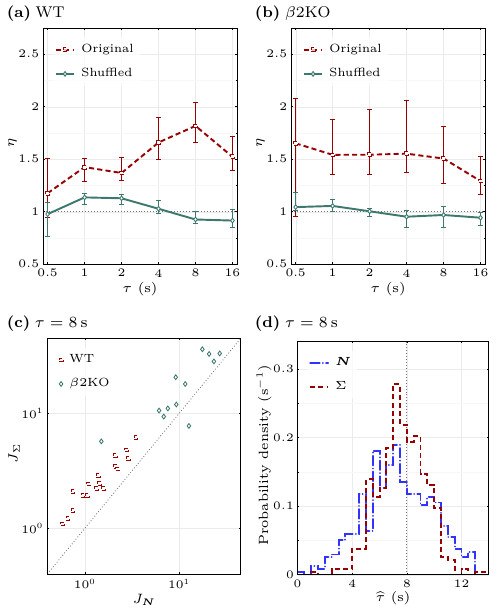}
  \caption{\textbf{Event order improves time estimation in retinal activity.} (a,b) Timing gain $\eta$ for the 17 WT and 12 $\beta2$KO recordings. Symbols denote the median across recordings, and error bars indicate the interquartile range. Red and green curves correspond to the original and randomized event order, respectively. The horizontal dotted line denotes $\eta = 1$. (c) Timing precision of the two clocks for all 29 recordings at $\tau = 8\,\mathrm{s}$. The diagonal denotes equal precision. (d) Distributions of the time estimates for the $\beta2$KO dorsal recording of June 6, 2008 (record 0), at $\tau = 8\,\mathrm{s}$. The vertical dotted line marks the actual duration.}
  \label{fig:retina_clock}
\end{figure}

\Cref{fig:retina_clock}(a,b) shows the median timing gain across recordings, with error bars indicating the interquartile range. For both WT and $\beta2$KO recordings, the median timing gain $\eta$ (red curves) lies above unity at all observation durations. Randomizing the event order brings the median timing gains down to approximately unity at all observation durations (green curves). The reduction in gain after randomization shows that the improvement depends on the original ordering.

As shown in \cref{fig:retina_clock}(c), both clocks generally achieve higher timing precision in the $\beta2$KO recordings than in the WT recordings. At $\tau = 8\,\mathrm{s}$, the median values of $J_{\boldsymbol{N}}$ and $J_\Sigma$ are $10.46$ and $14.99$ in $\beta2$KO, compared with $1.39$ and $2.43$ in WT. \Cref{fig:retina_clock}(d) shows the distributions of time estimates for one $\beta2$KO recording at the same duration. Retaining event order reduces the standard deviation from $2.48$ to $1.72\,\mathrm{s}$, corresponding to a $52.1\%$ reduction in timing variance and a gain of $\eta = 2.24$.

The timing gain also depends on the observation duration. In WT recordings, the median $\eta$ increases from $1.18$ at $0.5\,\mathrm{s}$ to $1.82$ at $8\,\mathrm{s}$, then decreases to $1.52$ at $16\,\mathrm{s}$. The gains in $\beta2$KO recordings vary less over the same range. These trends suggest that the usefulness of event order depends on the time scale over which collective activity develops. Counts cannot distinguish different recruitment patterns with the same total activity at each unit, whereas the order can help distinguish them. Whether the observation-duration dependence can be quantitatively related to a particular property of retinal waves remains an open question.

In a few recordings, adding event order gives $\eta < 1$ (see the outlier below the diagonal in \cref{fig:retina_clock}(c)). This can arise from variations in the dynamics across different segments of a recording. As shown in \cref{app:additional_results}, even when the sequence clock reduces timing fluctuations within individual segments, differences in the mean clock readings across segments can offset this improvement. Developing better optimization methods that are less sensitive to variations across recording segments remains an open problem.

\section{Recurrence in Cultured Neuronal Networks}
\label{sec:cultures}

We next examine spontaneous electrical activity in networks of rat hippocampal neurons cultured on a 4096-electrode array \cite{lonardoni2017recurrently}. The activity is characterized by periods of asynchronous firing interspersed with collective bursts that propagate through the network. As in the retinal experiments, we treat the spontaneous activity under each fixed experimental condition as autonomous stochastic dynamics. Here event labels identify recording electrodes rather than individual neurons. The spatial position of each electrode is determined by its row and column indices $(i_e,j_e)$ on the $64 \times 64$ array. We use $\boldsymbol{r}_e = ((i_e - 1)/64, (j_e - 1)/64)$, with $i_e,j_e = 1, \ldots, 64$. Of the 22 usable recording files from the public dataset \cite{lonardoni2017recurrently,lonardoni2018dataset}, two records labeled d2 show particularly large timing gains. They were obtained under a basal condition (before drug treatment) and under bicuculline (BIC), a drug that blocks GABA-A receptor-mediated inhibition. The clocks are calibrated at durations from $0.05$ to $1.6\,\mathrm{s}$.

\begin{figure}[t]
  \centering
  \includegraphics[width=\columnwidth]{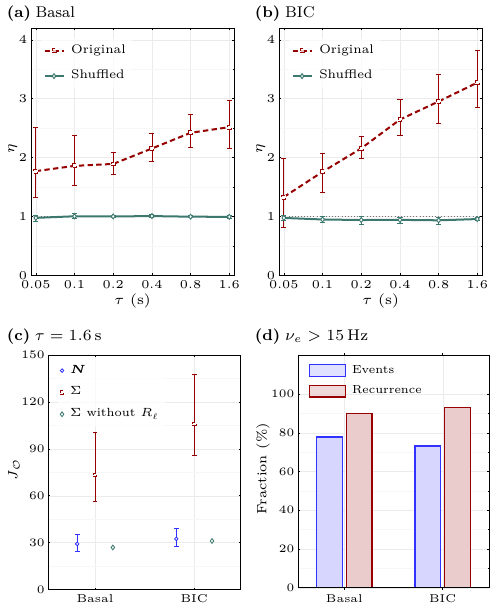}
  \caption{\textbf{Recurrence and timing precision in cultured neuronal networks.} (a,b) Timing gain of the d2 basal and BIC recordings, respectively, for the original and randomized event order. Error bars denote pointwise $95\%$ confidence intervals within each recording. (c) Timing precision at $\tau = 1.6\,\mathrm{s}$ for the count clock, the sequence clock, and the sequence clock with the eight recurrence terms removed while holding the other coefficients fixed. Error bars for the first two clocks denote pointwise $95\%$ confidence intervals; no new interval is calculated for the modified clock. (d) Fractions of all recorded events (blue) and of $\partial_\tau \langle Q_{\rm{rec}} \rangle$ at $\tau = 1.6\,\mathrm{s}$ (red) contributed by electrodes with $\nu_e > 15\,\mathrm{Hz}$ (see \cref{appeq:high_rate_fraction}).}
  \label{fig:culture_clock}
\end{figure}

At $1.6\,\mathrm{s}$, retaining event order increases the timing precision from $J_{\boldsymbol{N}} = 29.15$ to $J_\Sigma = 73.34$ in the basal condition and from $32.38$ to $105.95$ under BIC. The gains are $\eta = 2.52$ and $3.27$, with pointwise $95\%$ confidence intervals of $[2.16,2.98]$ and $[2.85,3.82]$, respectively. For both clocks, the relative error in the mean estimated duration is less than $0.6\%$. Retaining event order reduces the timing variance by $59.9\%$ and $70.1\%$ in the two recordings, respectively. As shown in \cref{fig:culture_clock}(a,b), the gain increases with the observation duration in both conditions, while randomizing order keeps it close to unity.

We then examine which properties of the event sequence contribute to this improvement. A particularly useful observable is the recurrence of an electrode label at a separation of $\ell$ events in the sequence,
\begin{equation}
  R_\ell(\Sigma_\tau) = \sum_{k=1}^{(n_\tau - \ell)_+} \mathbf{1}_{\{ e_k = e_{k+\ell} \}},
  \label{eq:recurrence}
\end{equation}
where $(x)_+ = \max(x,0)$. The sequence clock uses the excess recurrence relative to random order at fixed counts. Removing the eight recurrence terms ($\ell = 1, 2, 4, 8, 16, 32, 64, 128$) from the calibrated sequence clock, with all other coefficients fixed, reduces $J_\Sigma$ to $26.88$ in the basal condition and $31.02$ under BIC, close to the respective count-clock values. The improvement of these clocks is therefore largely associated with recurrence (\cref{fig:culture_clock}(c)). When only electrodes with mean event frequencies between $0.1$ and $15\,\mathrm{Hz}$ are retained, the timing gains decrease to $0.957$ and $0.959$ in the basal and BIC recordings, respectively (see Appendix C). To understand how event order suppresses timing fluctuations, we decompose the sequence readout into its count and order contributions (see \cref{apptab:count_order_variance}). In the basal recording, the negative covariance between these two contributions reduces the timing variance by $0.0491\,\mathrm{s}^2$ relative to the sum of their individual variances. Their combined variance is only $0.0350\,\mathrm{s}^2$, compared with $0.0574\,\mathrm{s}^2$ for the count contribution alone. Thus, the order contribution compensates for fluctuations in the count contribution.

The recurrence contribution is concentrated in high-frequency electrode activity. We define the mean event frequency of electrode $e$ as $\nu_e = n_e(T_{\rm{rec}})/T_{\rm{rec}}$, where $T_{\rm{rec}}$ is the duration of the complete recording and $n_e(T_{\rm{rec}})$ is the total number of events on that electrode. The 22 basal and 27 BIC electrodes with $\nu_e > 15\,\mathrm{Hz}$ contribute $77.7\%$ and $73.2\%$ of all recorded events, respectively. Writing $Q_{\rm{rec}}$ for the sum of the eight recurrence terms in the sequence clock, we find that these electrodes contribute $89.8\%$ and $92.9\%$ of $\partial_\tau \langle Q_{\rm{rec}} \rangle$ at $\tau = 1.6\,\mathrm{s}$ (\cref{fig:culture_clock}(d) and \cref{appeq:high_rate_fraction}). We interpret the high-frequency signals in these recordings as possible multiunit activity (MUA), in which a single electrode detects spikes from several nearby neurons \cite{rey2015past}. The high event frequency may thus reflect the aggregate firing of a local neuronal population. While the count clock measures the total activity of this population, the recurrence statistics capture how its activity is organized along the event sequence.

Event order plays an important role in inferring time irreversibility from non-Markovian observations \cite{blom2024milestoning,schwarz2026consistent}. Notably, $R_\ell$ is unchanged by reversing the entire sequence. The recurrence observable $R_\ell$ is invariant under reversal of the event sequence. Timing information need not encode an arrow of time: even a time-symmetric property of an event sequence can improve the precision with which its duration is estimated.

\section{Discussion and Conclusions}
\label{sec:discussion}

The information hierarchy of non-Markovian dynamics provides a physical basis for extracting timing information from experimental records. Our analysis of spontaneous retinal activity shows that the timing advantage of event order is a robust feature of the observed dynamics, persisting across a broad range of observation durations and under different genetic conditions. Our analysis of cultured neuronal networks reveals a specific dynamical structure that can give rise to this advantage: the recurrence of events on high-frequency electrodes. In particular, the order contribution compensates for fluctuations in the count contribution, allowing the same set of stochastic events to provide a more precise measure of time.

These findings suggest a route toward precise timekeeping in systems whose dynamical speeds fluctuate. Rather than suppressing fluctuations in the underlying dynamics, one may exploit the information encoded in their history to correct timing errors. This possibility raises several questions of broader interest. How does the organization of collective activity determine the timing information carried by event sequences? Can stochastic dynamics be designed to encode more timing information in their event order? Do biological systems exploit such information to coordinate processes in time? Addressing these questions may help establish a connection between the organization of stochastic dynamics and the physical principles of biological timekeeping.

\section{Acknowledgments}
\label{sec:acknowledgments}

GPT-6 Astra was used to assist with writing the data analysis code and polishing the manuscript. The authors take full responsibility for the work. This work is supported by the U.S. National Science Foundation under Grant No. DMR-2145256 and the Alfred P. Sloan Foundation Matter-to-Life Theory Award under Grant No. G-2025-25194.

\section{Data availability}

No new experiments were performed in this study. The retinal and cultured neuronal network recordings analyzed in this work are publicly available in Refs.~\cite{eglen2014dataset,lonardoni2018dataset}, respectively. The data analysis and plotting codes are openly available at \cite{data}.

\appendix

\crefalias{section}{appendix}
\crefalias{subsection}{appendix}

\section{Optimization of Time Estimators}
\label{app:optimization}

We consider a time estimator of the form
\begin{equation}
  \widehat{\tau} = b + \boldsymbol{a}^{\mathsf{T}} \boldsymbol{\phi},
\end{equation}
where $\boldsymbol{\phi}$ contains the observables calculated from an event record. The coefficients $\boldsymbol{a}$ and the offset $b$ are determined by calibration.

Let $\boldsymbol{m}_j$ and $C_j$ denote the sample mean and covariance matrix of the statistics calculated from calibration windows of duration $\tau_j$. The mean and variance of the time estimate are
\begin{subequations}
\begin{align}
  \langle \widehat{\tau} \rangle_j &= b + \boldsymbol{a}^{\mathsf{T}} \boldsymbol{m}_j, \\
  \Var_j[\widehat{\tau}] &= \boldsymbol{a}^{\mathsf{T}} C_j \boldsymbol{a}.
\end{align}
\end{subequations}
We minimize the average variance over $m$ calibration durations, subject to the condition that the mean readout equals the actual duration at each of them. Including a penalty on the coefficients, we obtain
\begin{subequations}
\label{appeq:optimization}
\begin{align}
  \min_{\boldsymbol{a},b} \quad& \boldsymbol{a}^{\mathsf{T}} H_\rho \boldsymbol{a}, \\
  \text{subject to} \quad& b + \boldsymbol{m}_j^{\mathsf{T}} \boldsymbol{a} = \tau_j, \qquad j = 1, \ldots, m,
\end{align}
\end{subequations}
where
\begin{equation}
  H_\rho = \overline{C}+\rho\diag(\overline{C}), \qquad \overline{C} = \frac{1}{m}\sum_{j=1}^{m}C_j.
\end{equation}
The term proportional to $\rho$ penalizes large coefficients after accounting for the scale of each statistic. It reduces the effect of statistical uncertainty in the estimated covariance matrix \cite{hoerl1970ridge}.

Eliminating the offset through $b = \tau_1 - \boldsymbol{m}_1^{\mathsf{T}} \boldsymbol{a}$ reduces the calibration conditions to
\begin{equation}
  D\boldsymbol{a} = \boldsymbol{d},
\end{equation}
where
\begin{equation}
  D= \begin{pmatrix}
    (\boldsymbol{m}_2 - \boldsymbol{m}_1)^{\mathsf{T}} \\
    \vdots \\
    (\boldsymbol{m}_m - \boldsymbol{m}_1)^{\mathsf{T}}
  \end{pmatrix}, \qquad
  \boldsymbol{d}= \begin{pmatrix}
    \tau_2 - \tau_1 \\
    \vdots \\
    \tau_m - \tau_1
  \end{pmatrix}.
\end{equation}
For $\rho > 0$, after removing statistics with vanishing variance, $H_\rho$ is positive definite. Provided that the calibration constraints are consistent, introducing Lagrange multipliers gives the optimal coefficients
\begin{subequations}
\begin{align}
  \boldsymbol{a}_* &= H_\rho^{-1} D^{\mathsf{T}} \left( DH_\rho^{-1} D^{\mathsf{T}} \right)^+ \boldsymbol{d},\\
  b_* &= \tau_1 - \boldsymbol{m}_1^{\mathsf{T}} \boldsymbol{a}_*,
\end{align}
\label{appeq:optimal_coefficients}
\end{subequations}
where the superscript ``$+$'' denotes the Moore-Penrose pseudoinverse. Statistics with vanishing variance in the records used for calibration are omitted before solving the quadratic problem.

The optimization thus requires only the means and covariances of the specified observables. Once the coefficients are determined, the same estimator is applied to records of different durations.

\section{Statistics and Data Analysis}
\label{app:features}

\subsection{Count and sequence statistics}

The count record $\boldsymbol{N}_\tau = (n_1, \ldots, n_M)$ contains the number of events observed in each channel. The sequence record $\Sigma_\tau = (e_1, \ldots, e_{n_\tau})$ additionally retains their order, where $n_\tau = \sum_e n_e$.

The count statistics include individual channel counts, nonlinear functions of the counts, and statistics describing how events are distributed among channels. These include the total event count, the number of active channels, and the concentration of events among channels. Spatial information is incorporated by grouping channels according to their positions and calculating the corresponding regional counts.

The additional sequence statistics describe three properties of event order:
\begin{itemize}
  \item[(i).] the occurrence of pairs of event labels at positions $k$ and $k+\ell$ for different values of $\ell$;
  \item[(ii).] the recurrence of the same event label and the spatial displacement between the two events;
  \item[(iii).] the distribution of event labels in successive portions of the sequence.
\end{itemize}

For example, the recurrence observable is
\begin{equation}
  R_\ell(\Sigma_\tau) = \sum_{k=1}^{(n_\tau - \ell)_+} \mathbf{1}_{\{ e_k = e_{k+\ell} \}},
  \label{appeq:recurrence}
\end{equation}
where $(x)_+ = \max(x,0)$. The spatial displacement observable is
\begin{equation}
  D_\ell(\Sigma_\tau) = \sum_{k=1}^{(n_\tau - \ell)_+} \left| \boldsymbol{r}_{e_{k+\ell}} - \boldsymbol{r}_{e_k} \right|^2,
\end{equation}
where $\boldsymbol{r}_e$ denotes the normalized spatial position associated with event label $e$, as defined in the main text. For retinal recordings, the coordinates are supplied by the \texttt{epos} field of the dataset and each nonconstant coordinate is rescaled independently to $[0,1]$. A constant coordinate is set to $1/2$. For cultured-network recordings, the coordinates follow from the electrode row and column indices. Here $\ell$ denotes the difference between two event indices, rather than the time interval between them.

To distinguish order structure from the effects of unequal channel counts, each sequence observable $S$ is centered relative to uniform random permutations of the event labels,
\begin{equation}
  \widetilde{S}(\Sigma_\tau) = S(\Sigma_\tau) - \mathbb{E}_{\rm{perm}} [S \,|\, \boldsymbol{N}_\tau].
  \label{appeq:centered_feature}
\end{equation}
For example,
\begin{equation}
  \mathbb{E}_{\rm{perm}}[R_\ell\,|\, \boldsymbol{N}_\tau] = \frac{(n_\tau - \ell)_+}{n_\tau (n_\tau - 1)} \sum_e n_e(n_e-1),
  \label{appeq:permutation_expectation}
\end{equation}
for $n_\tau \geq 2$, and is zero otherwise. The permutation expectation accounts for the recurrence expected solely from the channel counts. The count statistics include the permutation means of the pair, recurrence, and displacement statistics; for the retinal data they also include the means of the statistics describing successive portions of the sequence.

The sequence estimator uses the count statistics together with the additional centered sequence statistics. The numbers of statistics used by the two estimators are summarized in \cref{apptab:features}. All statistics are defined in the analysis code.

\begin{table}[t]
  \caption{Numbers of statistics used to construct the estimators in the two experimental systems. Here $M$ is the number of recorded units in a retinal recording.}
  \label{apptab:features}
  \begin{ruledtabular}
  \begin{tabular}{lcc}
    & Retina & Cultured network \\
    \midrule
    Count statistics & $3M+1052$ & $4934$ \\
    Order statistics & $534$ & $512$ \\
    Sequence total & $3M+1586$ & $5446$
  \end{tabular}
  \end{ruledtabular}
\end{table}

None of these statistics uses event timestamps, physical-time intervals, or the duration of the observation window.

\subsection{Determining the coefficients and measuring timing precision}
\label{app:validation}

Each experimental recording is divided into five consecutive segments, numbered from 0 to 4. When segment $k$ is used to measure timing precision, segment $(k+1)\bmod 5$ is used to choose $\rho$, and the other three segments are used to determine the estimator coefficients for each value of $\rho$. This procedure is repeated for all five segments.

The starting times of the observation windows are chosen independently of the recorded events. Windows containing no events are retained. We leave an unused interval at both ends of each segment, and no window crosses a segment boundary.

For each estimator, we consider the following values of $\rho$:
\begin{equation}
  \rho \in \{0.003,0.03,0.3,3,30,300,3000\}.
\end{equation}
For each value, we determine the estimator coefficients using the same three segments and calculate the mean squared difference between the estimated and actual durations in the segment used to choose $\rho$. We give each observation duration equal weight and choose the value of $\rho$ for which this error is smallest.

We then fix $\rho$ and the coefficients and estimate the durations of windows in the remaining segment, which was not used to determine them. We combine these estimates from all five segments, retaining negative values as well.

The mean clock rate is calculated using windows with the same starting points but slightly different durations,
\begin{equation}
  s_{\mathcal{O}}(\tau) \simeq \frac{ \langle \widehat{\tau}_{\mathcal{O}} \rangle_{(1+\delta)\tau} - \langle \widehat{\tau}_{\mathcal{O}} \rangle_{(1-\delta)\tau} }{2 \delta \tau}.
  \label{appeq:clock_rate}
\end{equation}
The window lengths are rounded to integer multiples of the time resolution of the recording, and the actual difference in duration is used in the calculation. The estimator coefficients remain fixed throughout this calculation.

The timing precision and timing gain are
\begin{equation}
  J_{\mathcal{O}}(\tau) = \frac{\tau^2 s_{\mathcal{O}}(\tau)^2}{\Var[\widehat{\tau}_{\mathcal{O}}]}, \qquad \eta(\tau) = \frac{J_\Sigma(\tau)}{J_{\boldsymbol{N}}(\tau)}.
\end{equation}

To obtain confidence intervals while accounting for temporal correlations, we resample groups of consecutive observations \cite{kunsch1989jackknife}. Specifically, we group windows within each segment according to their starting times, using time intervals of the lengths given in \cref{apptab:parameters}. We repeatedly draw these groups at random with replacement, retaining the original number of groups in each segment. For each repetition, we recalculate the timing precision and $\eta$ using the same groups for both estimators and for the three window durations used to calculate the mean clock rate. The estimator coefficients are kept fixed. The $95\%$ confidence interval is given by the 2.5th and 97.5th percentiles of the resulting values.

The main analysis parameters are summarized in \cref{apptab:parameters}. To examine the effect of event order, we randomly reorder the event labels within each window without changing any channel count. We then choose $\rho$ and determine the coefficients of the sequence estimator again, following the same procedure. The count estimator is unchanged. We also repeat the calculation of the mean clock rate using $\delta=0.1$ and check whether the results depend on the order assigned to events recorded at the same time within the resolution of the experiment.

\begin{table*}[t]
  \caption{Parameters used in the experimental data analysis. The cultured-network column refers to the two d2 recordings. For the remaining cultured-network recordings, confidence intervals are calculated using 500 repetitions.}
  \label{apptab:parameters}
  \begin{ruledtabular}
  \begin{tabular}{lcc}
    & Retina & Cultured network \\
    \midrule
    Calibration durations ($\mathrm{s}$)
    & $0.5,1,2,4,8,16$
    & $0.05,0.1,0.2,0.4,0.8,1.6$ \\
    Number of recording segments & $5$ & $5$ \\
    Unused interval at each segment end ($\mathrm{s}$)
    & $60$ & $9$ \\
    Maximum windows per segment and duration
    & $1200$ & $1200$ \\
    Relative change in duration $\delta$
    & $0.2$ & $0.2$ \\
    Time interval for grouping windows ($\mathrm{s}$)
    & $120$ & $30$ \\
    Number of repetitions
    & $1500$ & $1500$
  \end{tabular}
  \end{ruledtabular}
\end{table*}

\section{Detailed Data Analysis Results}
\label{app:additional_results}

\subsection{Timing gains in retinal activity}

\Cref{apptab:retinal_gains} summarizes the timing gain across all observation durations for the retinal data. The wild-type (WT) and $\beta2$ knockout ($\beta2$KO) groups contain 17 and 12 recordings, respectively. The improvement persists over a range of observation durations and is substantially reduced when the event order is randomized. Grouping the windows into intervals of $240\,\mathrm{s}$ instead of $120\,\mathrm{s}$ when calculating confidence intervals leaves the principal result unchanged: at both $\tau=4\,\mathrm{s}$ and $8\,\mathrm{s}$, 28 of the 29 recordings have confidence intervals for $\eta$ lying above unity.

\begin{table*}[t]
  \caption{Median timing gain $\eta$ across retinal recordings before and after randomizing event order. Here $n_+$ denotes the number of original recordings whose pointwise $95\%$ confidence interval for $\eta$ lies above unity.}
  \label{apptab:retinal_gains}
  \begin{ruledtabular}
  \begin{tabular}{ccccccc}
    & \multicolumn{3}{c}{WT}
    & \multicolumn{3}{c}{$\beta2$KO} \\
    $\tau$ ($\mathrm{s}$)
    & $\eta$ & Shuffled $\eta$ & $n_+$
    & $\eta$ & Shuffled $\eta$ & $n_+$ \\
    \midrule
    $0.5$ & $1.176$ & $0.976$ & $8$
          & $1.651$ & $1.043$ & $8$ \\
    $1$   & $1.424$ & $1.136$ & $13$
          & $1.541$ & $1.054$ & $10$ \\
    $2$   & $1.368$ & $1.127$ & $13$
          & $1.542$ & $1.003$ & $11$ \\
    $4$   & $1.659$ & $1.027$ & $17$
          & $1.552$ & $0.952$ & $11$ \\
    $8$   & $1.817$ & $0.925$ & $17$
          & $1.506$ & $0.968$ & $11$ \\
    $16$  & $1.523$ & $0.915$ & $16$
          & $1.291$ & $0.942$ & $7$
  \end{tabular}
  \end{ruledtabular}
\end{table*}

\subsection{Variation across recording periods}

To examine recordings in which the sequence estimator performs less well, we decompose the variance of the combined time estimates into fluctuations within each segment of the recording and differences between the mean readings of different segments.

Let $n_k$, $\overline{q}_k$, and $V_k$ denote the number, mean, and sample variance of the time estimates in segment $k$. The combined sample variance is
\begin{equation}
  V = V_{\rm{within}} + V_{\rm{between}},
\end{equation}
where
\begin{subequations}
\begin{align}
  V_{\rm{within}} &= \frac{1}{n-1}\sum_k(n_k-1)V_k,\\
  V_{\rm{between}} &= \frac{1}{n-1}\sum_k n_k(\overline{q}_k-\overline{q})^2,
\end{align}
\end{subequations}
with $n=\sum_k n_k$ and $\overline{q}=n^{-1}\sum_k n_k\overline{q}_k$.

\Cref{apptab:variance_decomposition} gives this decomposition for two retinal recordings with $\eta<1$ at $\tau=16\,\mathrm{s}$. In the first recording, the sequence estimator reduces the fluctuations within individual segments but increases the variation between their mean readings. In the second recording, both contributions increase.

\begin{table*}[t]
  \caption{Variance decomposition for two retinal recordings at $\tau=16\,\mathrm{s}$. R1 and R2 denote recordings \texttt{Beta2KO\_Nasal\_2008-06-28-0} and \texttt{Beta2KO\_Temporal\_2008-01-29-4}, respectively. All variance contributions are in units of $\mathrm{s}^2$.}
  \label{apptab:variance_decomposition}
  \begin{ruledtabular}
  \begin{tabular}{cccccc}
    Record
    & $V_{\boldsymbol{N},\rm{within}}$
    & $V_{\Sigma,\rm{within}}$
    & $V_{\boldsymbol{N},\rm{between}}$
    & $V_{\Sigma,\rm{between}}$
    & $\eta$ \\
    \midrule
    R1 & $12.284$ & $9.905$ & $1.868$ & $5.719$ & $0.968$ \\
    R2 & $9.326$ & $11.640$ & $1.486$ & $5.931$ & $0.622$
  \end{tabular}
  \end{ruledtabular}
\end{table*}

The contribution from differences between segments reflects both variations in the recorded dynamics and differences among the estimators whose coefficients were determined using different segments.

\subsection{Timing gains in cultured neuronal networks}

We analyze 22 usable recording files from the cultured-network dataset, excluding files with missing time-calibration information or duplicated recording intervals. The complete results are provided with the analysis code.

The largest gains are found in the two d2 recordings. At $\tau=1.6\,\mathrm{s}$, three of the remaining 20 recordings have confidence intervals for $\eta$ lying above unity. Their timing gains range from $0.584$ to $1.228$.

For the two d2 recordings, the sequence estimator can be decomposed into count and order contributions,
\begin{equation}
  \widehat{\tau}_\Sigma = C_\Sigma(\boldsymbol{N}_\tau) + Q_{\rm{ord}}(\Sigma_\tau),
\end{equation}
where
\begin{subequations}
\begin{align}
  C_\Sigma(\boldsymbol{N}_\tau) &= b_\Sigma + \boldsymbol{a}_{\Sigma,\boldsymbol{N}}^{\mathsf{T}} \boldsymbol{\phi}_{\boldsymbol{N}}(\boldsymbol{N}_\tau), \\
  Q_{\rm{ord}}(\Sigma_\tau) &= \boldsymbol{a}_{\Sigma,S}^{\mathsf{T}} \boldsymbol{\phi}_S(\Sigma_\tau).
\end{align}
\end{subequations}
Here $\boldsymbol{\phi}_S$ contains the additional centered order statistics, and $\boldsymbol{a}_{\Sigma,\boldsymbol{N}}$ and $\boldsymbol{a}_{\Sigma,S}$ are the coefficients of the count and order statistics in the sequence estimator, respectively. Both contributions are obtained from the same sequence estimator, with all coefficients determined together. The variance of the readout is
\begin{subequations}
\label{appeq:variance_components}
\begin{align}
  \Var[\widehat{\tau}_\Sigma] ={}& \Var[C_\Sigma] + \Var[Q_{\rm{ord}}]\\
  &+ 2 \Cov(C_\Sigma,Q_{\rm{ord}}).
\end{align}
\end{subequations}
The three contributions are given in \cref{apptab:count_order_variance}. The negative covariance shows that the order contribution compensates for fluctuations in the count contribution.

\begin{table*}[t]
  \caption{Variance decomposition of the sequence estimator for d2 at $\tau=1.6\,\mathrm{s}$. All entries are in units of $\mathrm{s}^2$.}
  \label{apptab:count_order_variance}
  \begin{ruledtabular}
  \begin{tabular}{lcccc}
    Condition
    & $\Var[C_\Sigma]$
    & $\Var[Q_{\rm{ord}}]$
    & $2\Cov(C_\Sigma,Q_{\rm{ord}})$
    & $\Var[\widehat{\tau}_\Sigma]$ \\
    \midrule
    Basal & $0.05741$ & $0.02666$ & $-0.04909$ & $0.03498$ \\
    BIC   & $0.04293$ & $0.01863$ & $-0.03753$ & $0.02403$
  \end{tabular}
  \end{ruledtabular}
\end{table*}

\subsection{Contribution of high-frequency activity}

To identify which sequence observables contribute to the timing gain, we examine the recurrence component of the calibrated estimator,
\begin{equation}
  Q_{\rm{rec}}(\Sigma_\tau) = \sum_{\ell\in\mathcal{L}} a_\ell\widetilde{R}_\ell(\Sigma_\tau),
\end{equation}
where $\mathcal{L}=\{1,2,4,8,16,32,64,128\}$.

Removing this component from the calibrated estimator while keeping all other coefficients fixed gives
\begin{equation}
  \widehat{\tau}_{\Sigma-\rm{rec}} = \widehat{\tau}_\Sigma-Q_{\rm{rec}}.
\end{equation}
We calculate its timing precision using the same windows as for the original estimator. The results are summarized in \cref{apptab:recurrence_contribution}.

\begin{table}[t]
  \caption{Timing precision before and after removing the recurrence contribution from the calibrated sequence estimator at $\tau=1.6\,\mathrm{s}$.}
  \label{apptab:recurrence_contribution}
  \begin{ruledtabular}
  \begin{tabular}{lccc}
    Condition & $J_{\boldsymbol{N}}$ & $J_\Sigma$
    & $J_{\Sigma-\rm{rec}}$\\
    \midrule
    Basal & $29.15$ & $73.34$ & $26.88$\\
    BIC   & $32.38$ & $105.95$ & $31.02$
  \end{tabular}
  \end{ruledtabular}
\end{table}

The recurrence observable can be decomposed into contributions from individual electrodes,
\begin{equation}
  R_\ell = \sum_e R_{\ell,e},
\end{equation}
where
\begin{equation}
  R_{\ell,e} = \sum_{k=1}^{(n_\tau - \ell)_+} \mathbf{1}_{\{e_k=e\}} \mathbf{1}_{\{e_{k+\ell}=e\}}.
\end{equation}
The corresponding centered contribution is
\begin{equation}
  \widetilde{R}_{\ell,e} = R_{\ell,e} - \frac{(n_\tau - \ell)_+ n_e(n_e-1)}{n_\tau (n_\tau - 1)},
\end{equation}
for $n_\tau \geq 2$, and is zero otherwise.

Writing
\begin{equation}
  Q_{{\rm{rec}},e} = \sum_{\ell \in \mathcal{L}} a_\ell \widetilde{R}_{\ell,e},
\end{equation}
we obtain the additive decomposition $Q_{\rm{rec}}=\sum_eQ_{{\rm{rec}},e}$. The fraction of the mean clock rate arising from recurrence that is contributed by electrodes with mean event rates $\nu_e > 15\,\mathrm{Hz}$ is therefore
\begin{equation}
  f_{\rm{high}} = \frac{ \displaystyle{\sum_{e:\nu_e>15\,\mathrm{Hz}}} \partial_\tau \langle Q_{{\rm{rec}},e}\rangle }{ \partial_\tau \langle Q_{\rm{rec}}\rangle },
  \label{appeq:high_rate_fraction}
\end{equation}
where the derivatives are calculated from the differences in mean readings at the two durations in \cref{appeq:clock_rate}.

\begin{table}[t]
  \caption{Contribution of electrodes with mean event rates above $15\,\mathrm{Hz}$. The clock-rate fraction refers to the contribution to $\partial_\tau \langle Q_{\rm{rec}}\rangle$.}
  \label{apptab:high_rate}
  \begin{ruledtabular}
  \begin{tabular}{lccc}
    Condition
    & Electrodes
    & Event fraction
    & Clock-rate fraction \\
    \midrule
    Basal & $22$ & $77.7\%$ & $89.8\%$ \\
    BIC   & $27$ & $73.2\%$ & $92.9\%$
  \end{tabular}
  \end{ruledtabular}
\end{table}

As shown in \cref{apptab:high_rate}, high-frequency electrodes account for approximately $90\%$ of the mean clock rate arising from recurrence in both conditions. Together with \cref{apptab:recurrence_contribution}, this identifies the recurrence structure of high-frequency activity as the principal contribution to the improvement of the sequence estimators.

As a complementary check, recalibrating both estimators using only channels with mean event rates between $0.1$ and $15\,\mathrm{Hz}$, the range used in the original study \cite{lonardoni2017recurrently}, gives $\eta=0.957$ and $0.959$ at $\tau=1.6\,\mathrm{s}$ in the basal and BIC conditions, respectively.

\bibliography{manuscript}

\end{document}